\documentclass[12pt]{article}
\usepackage{amssymb,amsfonts}
\usepackage{epsf,epsfig}
\def \la{\lambda}

\begin{document}
\baselineskip 18pt

\title{Low-temperature asymptotics for the transverse dynamical structure factor for a magnetically polarized $XX$ chain at small and negative frequencies}
\author{P.N. Bibikov}
\date{\it Russian State Hydrometeorological University, Saint-Petersburg, Russia}
%\affiliation{Saint-Petersburg State University}
\maketitle

\vskip5mm

\begin{abstract}
Using the truncated form factor expansion the low-temperature asymptotics for the transverse dynamical structure factor of the magnetically polarized
$XX$ chain is studied. Unlike the previous paper we do not use the representation of structure factor in terms of the corresponding magnetic susceptibility. This enables to
obtain correct results at small and negative frequencies.
\end{abstract}

\maketitle

\section{Introduction}

The present paper is a supplement of the previous one \cite{1} where the low-temperature asymptotics of the transverse dynamical structure factor (TDSF) was studied
for the magnetically polarized $XX$ chain \cite{2} related to the Hamiltonian
\begin{equation}
\hat H=-\frac{1}{2}\sum_{n=1}^N\Big[J\Big({\bf S}^+_n{\bf S}^-_{n+1}+{\bf S}^-_n{\bf S}^+_{n+1}\Big)+h\Big({\bf S}_n^z+{\bf S}_{n+1}^z-I\Big)\Big],
\end{equation}
where ${\bf S}_n^{\pm}$ and ${\bf S}_n^z$ is the standard triple of spin-$\frac{1}{2}$ operators associated with $n$-th site
\begin{equation}
[{\bf S}_m^+,{\bf S}_n^-]=2\delta_{m,n}{\bf S}_n^z,\qquad[{\bf S}_m^z,{\bf S}_n^{\pm}]=\pm\delta_{m,n}{\bf S}_n^{\pm},
\end{equation}
$I$ is the identity $2^N\times2^N$ matrix and $h$ is a magnetic field. As in \cite{1} we postulate here the periodic boundary conditions
\begin{equation}
{\bf S}_{N+1}\equiv{\bf S}_1.
\end{equation}

Hamiltonian (1) acts in the tensor product Hilbert space
\begin{equation}
{\cal H}=\prod_{n=1}^N\otimes{\cal V}_n,
\end{equation}
where each ${\cal V}_n$ is the copy of ${\mathbb C}^2$ generated by up $|\uparrow\rangle$ and down $|\downarrow\rangle$ polarized states and attached to the $n$-th site.

All the calculations in \cite{1} were based on the well known formula \cite{3,4}
\begin{equation}
S(\omega,q,T)=-\frac{1}{\pi(1-{\rm e}^{-\beta\omega})}{\rm Im}\chi(\omega,q,T),\qquad\omega\neq0,
\end{equation}
supplemented with the Dyson equation for the transverse magnetic susceptibility \cite{3,5,6,7}
\begin{equation}
\chi(\omega,q,T)=\frac{1}{\chi_0^{-1}(\omega,q)-\Sigma(\omega,q,T)}.
\end{equation}
Here $\chi_0(\omega,q)$ is the zero temperature susceptibility and $\Sigma(\omega,q,T)$ is the (thermally activated) magnon self-energy
\begin{equation}
\chi_0(\omega,q)=\chi(\omega,q,0)\Longleftrightarrow\Sigma(\omega,q,0)=0.
\end{equation}

The guiding idea of the calculations in \cite{1} was first suggested in \cite{4}. It asserts that in the gapped (massive) regime the low-temperature asymptotic of $S(\omega,q,T)$
completely depends only on the one- and two-magnon spectrums and hence may be obtained with the use of the so called truncated form factor expansion. It is well known however \cite{1,4,5} that a
direct derivation of this expansion for $S(\omega,q,T)$ (or $\chi(\omega,q,T)$) often yields a singular result for the line shape of the resonance contour even at $T>0$.
In order to avoid this pathology it was suggested in \cite{5} to search for the low temperature asymptotics not for $\chi(\omega,q,T)$ but for $\Sigma(\omega,q,T)$. The latter task however is not
simple if we work in the Matsubara temperature formalism where the Dyson equation (6) is usually proved by an analysis of the perturbation series. Namely in \cite{5} (6) was only postulated
and the expansion for $\Sigma(\omega,q,T)$ was obtained from the corresponding result for $\chi(\omega,q,T)$ with the use of the resummation procedure.
Contrary in \cite{1} the Dyson equation for the related to $\chi(\omega,q,T)$ real two time retarded Green function and the spectral representation for the corresponding
$\Sigma(\omega,q,T)$ were rigorously proved within the approach previously suggested by N. M. Plakida and Yu. A. Tserkovnikov \cite{6,7,8}.

Being rather successful near resonance the approach \cite{1,4,5} however fails at the vicinity $\omega=0$ and at $\omega<0$. Really according to (5) and the condition
\begin{equation}
S(\omega,q,T)\geq0,
\end{equation}
(which directly follows from the well known expansion (15) \cite{3}) ${\rm Im}\Sigma(\omega,q,T)$ should change its sign when $\omega$ passes throw 0.
This requirement however badly agrees with the form factor expansion for $\chi(\omega,q,T)$ usually based on the standard spectral decomposition \cite{1,3,4,5}
\begin{equation}
\chi(\omega,q,T)=\lim_{N\rightarrow\infty}\frac{1}{Z(T,N)}\sum_{\mu,\nu}\frac{{\rm e}^{-\beta E_{\nu}}-{\rm e}^{-\beta E_{\mu}}}{\omega+E_{\nu}-E_{\mu}+i\epsilon}
|\langle\nu|{\bf S}^+(q)|\mu\rangle|^2.
\end{equation}
Here $Z(T,N)$ is the partition function and
\begin{equation}
{\bf S}(q)\equiv\frac{1}{\sqrt{N}}\sum_{n=1}^N{\rm e}^{-iqn}{\bf S}_n,
\end{equation}
where according to the periodicity condition (3) it is implied
\begin{equation}
{\rm e}^{iqN}=1.
\end{equation}
Really the states related to the indices $\mu$ and $\nu$ in the sum (9) belong
to different magnon number sectors (for example if $\nu$ corresponds to the ground state then $\mu$ parameterizes the one-magnon states). But the $M$-th order form factor expansion implies
the cutoff of contributions from all $m$-magnon sectors with $m>M$. Hence if the index $\nu$ in (9) belongs to the $M$-magnon sector we should reduce  the corresponding term in (9) as follows
\begin{equation}
\frac{{\rm e}^{-\beta E_{\nu}}-{\rm e}^{-\beta E_{\mu}}}{\omega+E_{\nu}-E_{\mu}+i\epsilon}\longrightarrow\frac{{\rm e}^{-\beta E_{\nu}}}{\omega+E_{\nu}-E_{\mu}+i\epsilon}.
\end{equation}
But according to the well known formula
\begin{equation}
{\rm Im}\frac{1}{x+i\epsilon}=-\pi\delta(x),
\end{equation}
(12) yields the reduction
\begin{equation}
{\rm e}^{-\beta E_{\nu}}\Big(1-{\rm e}^{-\beta\omega}\Big)\delta(\omega+E_{\nu}-E_{\mu})\longrightarrow{\rm e}^{-\beta E_{\nu}}\delta(\omega+E_{\nu}-E_{\mu}),
\end{equation}
in ${\rm Im}\chi(\omega,q,T)$ under which the factor $1-{\rm e}^{-\beta\omega}$ in the denominator of (5) is not to be canceled! Being negligible near the resonance peak this error becomes
critical at $\omega\rightarrow0$ and at $\omega<0$ results in the wrong answer $S(\omega,q,T)<0$.

In order to obtain correct results for TDSF at the vicinity $\omega=0$ and at $\omega<0$ we suggest here the
truncated form factor expansion directly for $S(\omega,q,T)$ basing on the spectral representation \cite{3}
\begin{equation}
S(\omega,q,T)=\lim_{N\rightarrow\infty}\frac{1}{Z(T,N)}\sum_{\mu,\nu}{\rm e}^{-\beta E_{\nu}}|\langle\nu|{\bf S}^+(q)|\mu\rangle|^2\delta(\omega+E_{\nu}-E_{\mu}).
\end{equation}
Since this approach results in the resonance singularity at $\omega=E_{magn}(q)$ (here $E_{magn}(q)$ is the magnon energy, see (26)) \cite{1}
we shall decompose the TDSF on regular and singular components
\begin{equation}
S(\omega,q,T)=S^{(reg)}(\omega,q,T)+S^{(sing)}(\omega,q,T),
\end{equation}
implying
\begin{eqnarray}
&&S^{(reg)}(\omega,q,T)\neq\infty,\qquad\omega\neq E_{magn}(q),\nonumber\\
&&S^{(sing)}(\omega,q,T)=0,\qquad\omega\neq E_{magn}(q).
\end{eqnarray}
Just $S^{(reg)}(\omega,q,T)$ will be asserted as a reliable approximation for TDSF at small and negative $\omega$.

In order to avoid manipulations with delta-functions we additionally suggest an alternative method for evaluation of TDSF based on the formula
\begin{equation}
S(\omega,q,T)=-\frac{1}{\pi}{\rm Im}\xi(\omega,q,T),
\end{equation}
where the auxiliary quantity $\xi(\omega,q,T)$ is defined by the spectral decomposition
\begin{equation}
\xi(\omega,q,T)=\sum_{\mu,\nu}\frac{{\rm e}^{-\beta E_{\nu}}|\langle\nu|{\bf S}^+(q)|\mu\rangle|^2}{\omega+E_{\nu}-E_{\mu}+i\epsilon},
\end{equation}
(really (18) follows from (19), (15) and (13)).

As it will be shown in the paper the both approaches yield the same result for the low-temperature asymptotics of $S^{(reg)}(\omega,q,T)$.

\section{One- and two-magnon spectrums}

In the present paper we study only the case
\begin{equation}
h>|J|,
\end{equation}
under which all the terms in the sum (1) are non-negative operators and the system has the single zero-energy polarized ground state
\begin{equation}
|\emptyset\rangle=|\uparrow\rangle\otimes\dots\otimes|\uparrow\rangle,
\end{equation}
which is the tensor product of $N$ vectors $|\uparrow\rangle$.

The one-magnon sector is spanned on the states
\begin{equation}
|k\rangle=\frac{1}{\sqrt{N}}\sum_{n=1}^N{\rm e}^{ikn}|\downarrow_n\rangle,\qquad|\downarrow_n\rangle\equiv{\bf S}^-_n|\emptyset\rangle,
\end{equation}
where according to the periodicity condition (3)
\begin{equation}
{\rm e}^{ikN}=1.
\end{equation}
It may be readily proved that the system (22) is orthogonal and complete. Namely
\begin{eqnarray}
&&\sum_{j=0}^{N-1}|k_j\rangle\langle k_j|=\frac{1}{N}\sum_{j=0}^{N-1}\sum_{m,n=1}^N{\rm e}^{ik_j(m-n)}{\bf S}^-_m|\emptyset\rangle\langle\emptyset|{\bf S}^+_n
=\sum_{n=1}^N{\bf S}^-_n|\emptyset\rangle\langle\emptyset|{\bf S}^+_n,\\
&&\langle k_j|k_l\rangle=\frac{1}{N}\sum_{n=1}^N{\rm e}^{i(k_l-k_j)n}=\delta_{jl}.
\end{eqnarray}
The corresponding to $|k\rangle$ energy is
\begin{equation}
E_{magn}(k)=h-J\cos{k}.
\end{equation}
According to (26) and (20) the one-magnon sector really is gapped and
\begin{equation}
E_{gap}=E_{magn}(k_{gap})=h-|J|,
\end{equation}
where
\begin{eqnarray}
&&k_{gap}=0,\qquad J>0,\nonumber\\
&&k_{gap}=\pi,\qquad J<0.
\end{eqnarray}

A two-magnon state describes a pair of scattering magnons (there are not two-magnon bound states in the $XX$ chain \cite{1,2}) and has the form (see (35) in \cite{1})
\begin{equation}
|k,\kappa\rangle=\frac{2}{N}\sum_{n_1<n_2}{\rm e}^{ik(n_1+n_2)/2}\sin{\kappa(n_2-n_1)}|\downarrow_{n_1}\downarrow_{n_2}\rangle,\qquad
|\downarrow_{n_1}\downarrow_{n_2}\rangle\equiv{\bf S}^-_{n_1}{\bf S}^-_{n_2}|\emptyset\rangle,
\end{equation}
where $k$ satisfies (23) and without lost of generality one may put
\begin{equation}
0<\kappa<\pi.
\end{equation}
The corresponding energy is
\begin{equation}
E_{scatt}(k,\kappa)=2h-2J\cos{\frac{k}{2}}\cos{\kappa}.
\end{equation}

According to the periodicity (3)
\begin{eqnarray}
&&{\rm e}^{ik(n+N+1)/2}\sin{\kappa(N+1-n)}={\rm e}^{ik(1+n)/2}\sin{\kappa(n-1)},\qquad n=2,\dots,N,\nonumber\\
&&{\rm e}^{ik(0+n)/2}\sin{\kappa(n-0)}={\rm e}^{ik(n+N)/2}\sin{\kappa(N-n)},\qquad n=1,\dots,N-1.
\end{eqnarray}
Representing (32) in an equivalent form
\begin{eqnarray}
&&\Big({\rm e}^{i(k/2+\kappa)N}+1\Big){\rm e}^{i\kappa(1-n)}+\Big({\rm e}^{i(k/2-\kappa)N}+1\Big){\rm e}^{i\kappa(n-1)},\qquad n=2,\dots,N,\nonumber\\
&&\Big({\rm e}^{i(k/2+\kappa)N}+1\Big){\rm e}^{-i\kappa n}+\Big({\rm e}^{i(k/2-\kappa)N}+1\Big){\rm e}^{i\kappa n},\qquad n=1,\dots,N-1,
\end{eqnarray}
one readily reduce it to
\begin{equation}
{\rm e}^{i(k/2\pm\kappa)N}=-1.
\end{equation}

The solutions of (34) are slightly different for even and odd $N$. For $N=2M$ the total set of wave numbers is
\begin{eqnarray}
k_l=\frac{4\pi l}{N},\quad l=0,1,\dots M-1,\qquad
\kappa_{\lambda}=\frac{(2\lambda-1)\pi}{N},\quad\la=1,2,\dots M,\\
k_m=\frac{2(2m-1)\pi}{N},\quad m=1,2,\dots M,\qquad
\kappa_{\mu}=\frac{2\pi\mu}{N},\quad\mu=1,2,\dots M-1.
\end{eqnarray}
Since
\begin{equation}
{\rm e}^{ik_lN/2}=1,\qquad{\rm e}^{i\kappa_{\lambda}N}=-1,\qquad
{\rm e}^{ik_mN/2}=-1,\qquad{\rm e}^{i\kappa_{\mu}N}=1,
\end{equation}
(34) is really satisfied. Moreover as it is shown in the Appendix A the system of states (29), (35), (36) form a complete orthogonal basis of the
two-magnon sector. Namely
\begin{eqnarray}
&&\sum_{l,\lambda}|k_l,\kappa_{\lambda}\rangle\langle k_l,\kappa_{\lambda}|+\sum_{m,\mu}|k_m,\kappa_{\mu}\rangle\langle k_m,\kappa_{\mu}|=\sum_{1\leq n_1<n_2\leq N}
{\bf S}^-_{n_1}{\bf S}^-_{n_2}|\emptyset\rangle\langle\emptyset|{\bf S}^+_{n_1}{\bf S}^+_{n_2},\\
&&\langle k,\kappa|\tilde k,\tilde\kappa\rangle=\delta_{k\tilde k}\delta_{\kappa\tilde\kappa}.
\end{eqnarray}

\section{Direct evaluation of $S_1^{(reg)}(\omega,q,T)$}

The truncated form factor expansion is based on the decomposition of the total Hilbert space (4) into the direct sum of the $m$-magnon sectors
\begin{equation}
{\cal H}=\oplus_{m=0}^N{\cal H}_m\qquad\hat Q\Big|_{{\cal H}_m}=m
\end{equation}
This follows from the commutativity of the Hamiltonian (1) with the magnon number operator
\begin{equation}
\hat Q=\sum_n\Big(\frac{1}{2}-{\bf S}_n^z\Big).
\end{equation}
The one-dimensional subspace ${\cal H}_0$ is spanned on $|\emptyset\rangle$. The one-and two-magnon sectors are spanned on (22) and (29).
Decomposition (40) results in the following expansions
\begin{equation}
Z(T,N)=1+\sum_{m=1}^NZ_m(T,N),\qquad{\cal J}(\omega,q,T,N)=\sum_{m=0}^{N-1}{\cal J}_m(\omega,q,T,N),
\end{equation}
for the partition sum $Z_m(T,N)$ and an auxiliary one
\begin{equation}
{\cal J}(\omega,q,T,N)\equiv\sum_{\mu,\nu}{\rm e}^{-\beta E_{\nu}}|\langle\nu|{\bf S}^+(q)|\mu\rangle|^2\delta(\omega+E_{\nu}-E_{\mu}).
\end{equation}
Here in (42)
\begin{eqnarray}
&&{\cal J}_m(\omega,q,T,N)\equiv\sum_{\mu,\nu}{\rm e}^{-\beta E_{\nu}}
|\langle\nu|{\bf S}^+(q)|\mu\rangle|^2\delta(\omega+E_{\nu}-E_{\mu}),\qquad|\nu\rangle\in{\cal H}_m,\quad|\mu\rangle\in{\cal H}_{m+1},\nonumber\\
&&Z_m(T,N)\equiv\sum_{\nu}{\rm e}^{-\beta E_{\nu}},\qquad|\nu\rangle\in{\cal H}_m.
\end{eqnarray}
From (42) follows the form factor expansion for the TDSF (15)
\begin{equation}
S(\omega,q,T)=\sum_{m=0}^{\infty}S_m(\omega,q,T),\qquad S_m(\omega,q,T)=O\Big({\rm e}^{-m\beta E_{gap}}\Big),
\end{equation}
where
\begin{eqnarray}
&&S_0(\omega,q,T)=\lim_{N\rightarrow\infty}{\cal J}_0(\omega,q,T,N),\nonumber\\
&&S_1(\omega,q,T)=\lim_{N\rightarrow\infty}({\cal J}_1(\omega,q,T,N)-Z_1(T,N){\cal J}_0(\omega,q,T,N)).
\end{eqnarray}
According to (44)
\begin{eqnarray}
&&Z_1(T,N)=\sum_k{\rm e}^{-\beta E_{magn}(k)},\nonumber\\
&&{\cal J}_0(\omega,q,T,N)=\sum_k|\langle\emptyset|{\bf S}^+(q)|k\rangle|^2\delta(\omega-E_{magn}(k))=\delta(\omega-E_{magn}(q)),\nonumber\\
&&{\cal J}_0(\omega,q,T,N)=\sum_{k,\kappa}{\rm e}^{-\beta E_{magn}(k-q)}|\langle k-q|{\bf S}^+(q)|k,\kappa\rangle|^2\nonumber\\
&&\cdot\delta(\omega+E_{magn}(k-q)-E_{scatt}(k,\kappa)),
\end{eqnarray}
and a substitution of (47) into (46) yields
\begin{eqnarray}
&&S_0(\omega,q,T)=\delta(\omega-E_{magn}(q)),\nonumber\\
&&S_1(\omega,q,T)=\lim_{N\rightarrow\infty}\sum_k{\rm e}^{-\beta E_{magn}(k-q)}\Big(\sum_{\kappa}|\langle k-q|{\bf S}^+(q)|k,\kappa\rangle|^2\nonumber\\
&&\cdot\delta(\omega+E_{magn}(k-q)-E_{scatt}(k,\kappa))-\delta(\omega-E_{magn}(q))\Big).
\end{eqnarray}

As it is shown in the appendix B
\begin{equation}
\langle k-q|{\bf S}^+(q)|k,\kappa\rangle=\frac{2i\sin{\kappa}}{N(\cos{\alpha}-\cos{\kappa})},
\end{equation}
where
\begin{equation}
\alpha=\frac{k}{2}-q.
\end{equation}
At the same time it may be readily checked according to (26), (31) and (50) that
\begin{equation}
E_{magn}(k-q)-E_{scatt}(k,\kappa)=-E_{magn}(q)+2J\cos{\frac{k}{2}}(\cos{\kappa}-\cos{\alpha}).
\end{equation}
So the condition
\begin{equation}
\omega+E_{magn}(k-q)-E_{scatt}(k,\kappa)=0,
\end{equation}
yields
\begin{equation}
\cos{\alpha}-\cos{\kappa}=\frac{\omega-E_{magn}(q)}{2J\cos{k/2}}.
\end{equation}
Correspondingly
\begin{equation}
\sin{\kappa}=\frac{\sqrt{-D(k,\omega,q)}}{2|J|\cos{k/2}},
\end{equation}
where
\begin{equation}
D(k,\omega,q)=(\omega-E_{magn}(q)-2J\cos{\frac{k}{2}}\cos{\alpha})^2-4J^2\cos^2{\frac{k}{2}}.
\end{equation}

It may be readily proved from (51) and (54) that
\begin{equation}
0<\sin{\kappa}<1,\quad-1<\cos{\kappa}<1\quad\Longleftrightarrow\quad D(k,\omega,q)<0.
\end{equation}
In order to study this conditions it is convenient to rewrite (55) in the form
\begin{equation}
D(k,\omega,q)=(\omega-\Phi_{down}(q,k))(\omega-\Phi_{up}(q,k)),
\end{equation}
where
\begin{eqnarray}
&&\Phi_{down}(q,k)=h+J\cos{(k-q)}-2|J|\cos{\frac{k}{2}},\nonumber\\
&&\Phi_{up}(q,k)=h+J\cos{(k-q)}+2|J|\cos{\frac{k}{2}}.
\end{eqnarray}
According to (57) the condition (56) reduces to
\begin{equation}
D(k,\omega,q)<0\Longleftrightarrow\Phi_{down}(q,k)<\omega<\Phi_{up}(q,k).
\end{equation}

A substitution of (53) and (54) into (49) gives with the use of (51)
\begin{eqnarray}
&&|\langle k-q|{\bf S}^+(q)|k,\kappa\rangle|^2\delta(\omega+E_{magn}(k-q)-E_{scatt}(k,\kappa))\nonumber\\
&&=\frac{8|J|\sqrt{-D(k,\omega,q)}\sin{\kappa}}{N^2(\omega-E_{magn}(q))^2}\cos{\frac{k}{2}}\delta(\omega-E_{magn}(q)+2J\cos{\frac{k}{2}}(\cos{\kappa}-\cos{\alpha})).
\end{eqnarray}
Excluding now the singular term (proportional to $\delta(\omega-E_{magn}(q))$) and using the standard substitutions
\begin{equation}
\sum_k\longrightarrow\frac{N}{2\pi}\int dk,\qquad\sum_{\kappa}\longrightarrow\frac{N}{2\pi}\int d\kappa,
\end{equation}
one readily gets from (48) and (60)
\begin{equation}
S_1^{(reg)}(\omega,q,T)=-\frac{{\rm Im}\Sigma_1(\omega,q,T)}{\pi(\omega-E_{magn}(q))^2},\qquad\omega\neq E_{magn}(q).
\end{equation}
Here
\begin{equation}
\Sigma_1(\omega,q,T)=-\frac{i}{\pi}\int_{-\pi}^{\pi}dk{\rm e}^{-\beta E_{magn}(k-q)}\sqrt{|D(k,\omega,q)|}
\Theta(\Phi_{up}(q,k)-\omega)\Theta(\omega-\Phi_{down}(q,k)),
\end{equation}
is the first term of the cluster expansion for the magnon self-energy obtained in \cite{1}.

\section{Alternative evaluation of $S_1(\omega,q,T)$}

In the same manner as in (48) we may readily get the cluster expansion for $\xi(\omega,q,T)$
\begin{equation}
\xi(\omega,q,T)=\xi_0(\omega,q)+\sum_{m=1}^{\infty}\xi_m(\omega,q,T),\qquad\xi_m(\omega,q,T)=O\Big({\rm e}^{-m\beta E_{gap}}\Big),
\end{equation}
where
\begin{eqnarray}
&&\xi_0(\omega,q)=\lim_{N\rightarrow\infty}\sum_k\frac{|\langle\emptyset|{\bf S}^+(q)|k\rangle|^2}{\omega-E_{magn}(k)+i\epsilon}
=\frac{1}{\omega-E_{magn}(q)+i\epsilon},\nonumber\\
&&\xi_1(\omega,q,T)=\lim_{N\rightarrow\infty}\sum_{k}{\rm e}^{-\beta E_{magn}(k-q)}\Big(\sum_{\kappa}\frac{|\langle k-q|{\bf S}^+(q)|k,\kappa\rangle|^2}{\omega+E_{magn}(k-q)-E_{scatt}(k,\kappa)+i\epsilon}\nonumber\\
&&-\frac{1}{\omega-E_{magn}(q)+i\epsilon}\Big).
\end{eqnarray}

Since there are two types of two-magnon states (35) and (36) we have the decomposition
\begin{equation}
\xi_1(\omega,q,T)=\xi_1^{(1)}(\omega,q,T)+\xi_1^{(2)}(\omega,q,T),
\end{equation}
where
\begin{eqnarray}
&&\xi_1^{(1)}(\omega,q,T)=\lim_{N\rightarrow\infty}\sum_{l=0}^{M-1}{\rm e}^{-\beta E_{magn}(k_l-q)}\Big(\sum_{\lambda=1}^M\frac{|\langle k_l-q|{\bf S}^+(q)|k_l,\kappa_{\lambda}\rangle|^2}{\omega+E_{magn}(k_l-q)-E_{scatt}(k_l,\kappa_{\lambda})+i\epsilon}\nonumber\\
&&-\frac{1}{\omega-E_{magn}(q)+i\epsilon}\Big),\nonumber\\
&&\xi_1^{(2)}(\omega,q,T)=\lim_{N\rightarrow\infty}\sum_{m=0}^M{\rm e}^{-\beta E_{magn}(k_m-q)}\Big(\sum_{\mu=1}^{M-1}\frac{|\langle k_m-q|{\bf S}^+(q)|k_m,\kappa_{\mu}\rangle|^2}{\omega+E_{magn}(k_m-q)-E_{scatt}(k_m,\kappa_{\mu})+i\epsilon}\nonumber\\
&&-\frac{1}{\omega-E_{magn}(q)+i\epsilon}\Big)
\end{eqnarray}

Introducing the new variables
\begin{equation}
x=\cos{\kappa},\qquad a=J\cos{\frac{k}{2}},\qquad b=\omega-2h+E_{magn}(k-q,h),\qquad c=\cos{\alpha},
\end{equation}
we readily obtain from (31) and (49) the following compact representation
\begin{equation}
\frac{|\langle k-q|{\bf S}^+(q)|k,\kappa\rangle|^2}{\omega+E_{magn}(k-q)-E_{scatt}(k,\kappa)+i\epsilon}=\frac{4(1-x^2)}{N^2(x-c)^2(2ax+b+i\epsilon)}.
\end{equation}

Using the identity
\begin{eqnarray}
&&\frac{1-x^2}{(x-c)^2(2ax+b+i\epsilon)}=\frac{1}{(2ac+b+i\epsilon)}\Big(\frac{1-c^2}{(x-c)^2}-\frac{2c}{x-c}\Big)\nonumber\\
&&-\frac{1}{(2ac+b+i\epsilon)^2}\Big(\frac{2a(1-c^2)}{x-c}+\frac{(b+i\epsilon)^2-4a^2}{2ax+b+i\epsilon}\Big),
\end{eqnarray}
and taking into account that
\begin{equation}
2ac+b+i\epsilon=\omega-E_{magn}(q)+i\epsilon,\qquad (b+i\epsilon)^2-4a^2=D(k,\omega+i\epsilon,q),
\end{equation}
we readily get the following expansion for (69)
\begin{eqnarray}
\frac{|\langle k-q|{\bf S}^+(q)|k,\kappa\rangle|^2}{\omega+E_{magn}(k-q)-E_{scatt}(k,\kappa)+i\epsilon}=\frac{4}{N^2}\Big[\Big(\frac{\sin^2\alpha}{(\cos{\kappa}-\cos{\alpha})^2}-
\frac{2\cos{\alpha}}{\cos{\kappa}-\cos{\alpha}}\Big)\nonumber\\
\cdot\frac{1}{\omega-E_{magn}(q)+i\epsilon}-\Big(\frac{2a\sin^2{\alpha}}{\cos{\kappa}-\cos{\alpha}}+\frac{D(k,\omega+i\epsilon,q)}{2ax+b+i\epsilon}\Big)\frac{1}{(\omega-E_{magn}(q)+i\epsilon)^2}\Big].
\end{eqnarray}

As it is shown in Appendix B
\begin{eqnarray}
&&\frac{4}{N^2}\sum_{\lambda=1}^M\Big(\frac{\sin^2{\alpha_l}}{(\cos{\kappa_{\lambda}}-\cos{\alpha_l})^2}-\frac{2\cos{\alpha_l}}{\cos{\kappa_{\lambda}}-\cos{\alpha_l}}\Big)=1+\delta_{\alpha_l,0}
+\delta_{|\alpha_l|,\pi},\nonumber\\
&&\frac{4}{N^2}\sum_{\mu=1}^{M-1}\Big(\frac{\sin^2{\alpha_m}}{(\cos{\kappa_{\mu}}-\cos{\alpha_m})^2}-\frac{2\cos{\alpha_m}}{\cos{\kappa_{\mu}}-\cos{\alpha_m}}\Big)=1-\frac{4}{N^2},\nonumber\\
&&\frac{4}{N^2}\sum_{\lambda=1}^M\frac{\sin^2{\alpha_l}}{\cos{\kappa_{\lambda}}-\cos{\alpha_l}}=0,\nonumber\\
&&\frac{4}{N^2}\sum_{\mu=1}^{M-1}\frac{\sin^2{\alpha_m}}{\cos{\kappa_{\mu}}-\cos{\alpha_m}}=-\frac{4\cos{\alpha_m}}{N^2},
\end{eqnarray}
where following (50) we introduced notations
\begin{equation}
\alpha_l\equiv\frac{k_l}{2}-q,\qquad \alpha_m\equiv\frac{k_m}{2}-q.
\end{equation}

A substitution of (72) into (67) gives with the use of (73)
\begin{eqnarray}
&&\xi_1^{(1)}(\omega,q,T)=-\lim_{N\rightarrow\infty}\frac{4}{(\omega-E_{magn}(q)+i\epsilon)^2N^2}\sum_{k_l,\kappa_{\lambda}}
\frac{{\rm e}^{-\beta E_{magn}(k_l-q)}D(k_l,\omega+i\epsilon,q)}{2a\cos{\kappa_{\lambda}}+b+i\epsilon}\nonumber\\
&&+\frac{{\rm e}^{-\beta E_{magn}(q)}}{\omega-E_{magn}(q)+i\epsilon}\nonumber\\
&&\xi_1^{(2)}(\omega,q,T)=-\lim_{N\rightarrow\infty}\frac{4}{(\omega-E_{magn}(q)+i\epsilon)^2N^2}
\sum_{k_m,\kappa_{\mu}}
\frac{{\rm e}^{-\beta E_{magn}(k_m-q)}D(k_m,\omega+i\epsilon,q)}{2a\cos{\kappa_{\mu}}+b+i\epsilon}.\qquad
\end{eqnarray}
Using now the $N\rightarrow\infty$ substitutions
\begin{equation}
\frac{1}{N}\sum_{k_l},\,\frac{1}{N}\sum_{k_m}\longrightarrow\frac{1}{4\pi}\int_{0}^{2\pi}dk,\qquad\frac{1}{N}\sum_{\kappa_{\lambda}},\,\frac{1}{N}\sum_{\kappa_{\mu}}
\longrightarrow\frac{1}{2\pi}\int_{0}^{\pi}d\kappa,
\end{equation}
and formula (66) one readily gets from (75)
\begin{eqnarray}
&&\xi_1(\omega,q,T)=\frac{1}{\pi(\omega-E_{magn}(q)+i\epsilon)^2}\int_{0}^{2\pi}dk{\rm e}^{-\beta E_{magn}(k-q)}\tilde\Gamma(k,\omega,q)\nonumber\\
&&+\frac{{\rm e}^{-\beta E_{magn}(q)}}{\omega-E_{magn}(q)+i\epsilon},
\end{eqnarray}
where
\begin{equation}
\tilde\Gamma(k,\omega,q)=\frac{1}{\pi}\int_0^{\pi}\frac{D(k,\omega+i\epsilon,q)d\kappa}{2a\cos{\kappa}+b+i\epsilon}=\frac{1}{2\pi i}\oint_{|z|=1}dz
\frac{4a^2-(b+i\epsilon)^2}{a(z^2+1)+(b+i\epsilon)z},
\end{equation}
is the same as in equation (94) of \cite{1}. So (77) and equation (93) in \cite{1} yield
\begin{equation}
\xi_1(\omega,q,T)=\frac{\Sigma_1(\omega,q,T)}{(\omega-E_{magn}(q)+i\epsilon)^2}+\frac{{\rm e}^{-\beta E_{magn}(q)}}{\omega-E_{magn}(q)+i\epsilon}.
\end{equation}
Now a substitution of (79) into (18) gives exactly the result (62) for $S^{(reg)}_1(\omega,q,T)$.

\section{Comparison with the TDSF obtained on the ground of the Dyson equation}

It is instructive to compare (62) with the result obtained in \cite{1} on the ground of the Dyson equation
\begin{equation}
S_1^{(DY)}(\omega,q,T)=-\frac{1}{\pi(1-{\rm e}^{-\beta\omega})}\cdot\frac{{\rm Im}\Sigma_1(\omega,q,T)}{(\omega-E_{magn}(q)-{\rm Re}\Sigma_1(\omega,q,T))^2+({\rm Im}\Sigma_1(\omega,q,T))^2}.
\end{equation}

The two expressions (62) and (80) has the two main differences. First of all (80) contains the denominator $1-{\rm e}^{-\beta\omega}$ while (62) does not. This subject was already discussed in the Introduction.
Second (80) includes ${\rm Re}\Sigma_1(\omega,q,T)$ while (62) does not. This disagreement becomes clear if we notice that (80) follows from (5), (6) and the relation
\begin{equation}
\chi_0^{-1}(\omega,q)=\omega-E_{magn}(q)+i\epsilon.
\end{equation}
Hence ${\rm Im}\chi_1(\omega,q,T)$ still contains ${\rm Re}\Sigma_1(\omega,q,T)$ in the denominator. At the same time we can not transfer $\Sigma_1(\omega,q,T)$
into the denominator of $\xi_0(\omega,q)+\xi_1(\omega,q,T)$ postulating for example the "resummation procedure"
\begin{equation}
\frac{1}{\omega-E_{magn}(q)+i\epsilon}+\frac{\Sigma_1(\omega,q,T)}{(\omega-E_{magn}(q))^2}\longrightarrow\frac{1}{\omega-E_{magn}(q)-\Sigma_1(\omega,q,T)},
\end{equation}
because $\xi(\omega,q,T)$ a priori does not satisfy something like the Dyson equation. That is why ${\rm Re}\Sigma_1(\omega,q,T)$ cancels when we evaluate ${\rm Im}\xi_1(\omega,q,T)$.

What formula is more correct (62) or (80)? Obviously $S_1^{(reg)}(\omega,q,T)$ turns to infinity at $\omega\rightarrow E_{magn}(q)$ while $S_1^{(DY)}(\omega,q,T)$ may be singular at $\omega=0$ if
\begin{equation}
\omega_{min}(q)=h-3|J|\cos{\frac{|q|+k_{gap}-\pi}{3}}<0,
\end{equation}
(the latter condition guarantees that $S(0,q,T)\neq0$ \cite{1}).
Hence it seems natural to introduce the intermediate frequency $0<\omega_s(q,T)<E_{magn}(q)$ defined by the condition
\begin{equation}
S_1^{(reg)}(\omega_s(q,T),q,T)=S_1^{(DY)}(\omega_s(q,T),q,T).
\end{equation}
At $\omega<\omega_s(q,T)$ one should use the formula (62) while at $\omega>\omega_s(q,T)$ the formula (80).

\section{Summary and discussion}

In the present paper we evaluated the low-temperature asymptotic for TDSF of the magnetically polarized $XX$ chain directly from the definition (15). We also confirmed the result by
alternative calculations according to the formula (18) with the use of an auxiliary quantity $\xi(\omega,q,T)$ (19). We assert that the obtained formula (62) adequately describes TDSF at very
small and negative frequencies but becomes completely incorrect near the resonance. According to this result supplemented by the result of the paper \cite{1} we have introduced
the frequency $\omega_s(q,T)$ which separates between the small and resonance frequency regions. In the former one the TDSF is described by the formula (62) suggested in the present paper
while in the latter by the formula (80) obtained previously \cite{1}.

Also from (62) follows that at the $O({\rm e}^{-\beta E_{gap}})$ level the magnetically polarized $XX$ chain does not possess a zero-frequency singularity in the TDSF \cite{9}.
So its isothermal and isolated transverse susceptibilities (on this level) coincide.

\appendix
\renewcommand{\theequation}{\thesection.\arabic{equation}}

\section{Orthogonality and completeness of the two-magnon basis}
\setcounter{equation}{0}
Let us represent a two-magnon state in the form
\begin{equation}
|k,\kappa\rangle=\frac{2}{\sqrt{N}}\sum_{n_1<n_2}{\rm e}^{ik(n_1+n_2)/2}\varphi_{n_2-n_1}(\kappa)|\downarrow_{n_1}\downarrow_{n_2}\rangle,
\end{equation}
where
\begin{equation}
\varphi_n(\kappa)=\bar\varphi_n(\kappa)=\frac{2\sin{\kappa n}}{\sqrt{N}}.
\end{equation}
Periodicity condition (34) may be represented now in an alternative form
\begin{equation}
\varphi_{N-n}(\kappa)={\rm e}^{\pm ikN/2}\varphi_n(\kappa).
\end{equation}

According to (A.1) and (A.3) one has
\begin{eqnarray}
&&\langle k,\kappa|\tilde k,\tilde\kappa\rangle=\frac{1}{N}\sum_{1\leq n_1<n_2\leq N}{\rm e}^{i(\tilde k-k)(n_1+n_2)/2}
\bar\varphi_{n_2-n_1}(\kappa)\varphi_{n_2-n_1}(\tilde\kappa)\nonumber\\
&&=\frac{1}{N}\sum_{n=1}^{N-1}\bar\varphi_n(\kappa)\varphi_n(\tilde\kappa)\sum_{m=n/2+1}^{N-n/2}{\rm e}^{i(\tilde k-k)m}
=\frac{1}{N}\sum_{n=1}^{N-1}\bar\varphi_{N-n}(\kappa)\varphi_{N-n}(\tilde\kappa)\nonumber\\
&&\cdot\sum_{m=n/2+1}^{N-n/2}{\rm e}^{i(\tilde k-k)(m+N/2)}=\frac{1}{N}\sum_{n=1}^{N-1}\bar\varphi_n(\kappa)\varphi_n(\tilde\kappa)
\sum_{m=(N-n)/2+1}^{N-(N-n)/2}{\rm e}^{i(\tilde k-k)(m+N/2)}\nonumber\\
&&=\frac{1}{N}\sum_{n=1}^{N-1}\bar\varphi_n(\kappa)\varphi_n(\tilde\kappa)\sum_{m=N-n/2+1}^{N+n/2}{\rm e}^{i(\tilde k-k)m}.
\end{eqnarray}
Now from (A.4) readily follows
\begin{equation}
\langle k,\kappa|\tilde k,\tilde\kappa\rangle=\frac{1}{2N}\sum_{n=1}^{N-1}\bar\varphi_n(\kappa)\varphi_n(\tilde\kappa)\sum_{m=1+n/2}^{N+n/2}{\rm e}^{i(\tilde k-k)m}
=\frac{\delta_{k,\tilde k}}{2}\sum_{n=1}^{N-1}\bar\varphi_n(\kappa)\varphi_n(\tilde\kappa).
\end{equation}
But (A.2) yields
\begin{equation}
\frac{1}{2}\sum_{n=1}^{N-1}\bar\varphi_n(\kappa)\varphi_n(\tilde\kappa)=\delta_{\kappa,\tilde\kappa}-\delta_{\kappa,-\tilde\kappa}=\delta_{\kappa,\tilde\kappa}.
\end{equation}
A combination of (A.5) and (A.6) gives (39).

The completeness condition
\begin{equation}
\sum_{k,\kappa}|k,\kappa\rangle\langle k,\kappa|=\sum_{n_1<n_2}{\bf S}^-_{n_1}{\bf S}^-_{n_2}|\emptyset\rangle\langle\emptyset|{\bf S}^+_{n_1}{\bf S}^+_{n_2},
\end{equation}
is equivalent to the formula
\begin{equation}
W\equiv\frac{4}{N^2}\sum_{k,\kappa}{\rm e}^{ik(n_1+n_2-\tilde n_1-\tilde n_2)/2}\sin{\kappa(n_2-n_1)}\sin{\kappa(\tilde n_2-\tilde n_1)}=\delta_{n_1\tilde n_1}\delta_{n_2\tilde n_2},
\end{equation}
where it is implied that
\begin{equation}
n_2>n_1,\qquad\tilde n_2>\tilde n_1.
\end{equation}
It is convenient to pass in (A.8) from $k$ and $\kappa$ to the individual magnon wave numbers
\begin{equation}
k_1=\frac{k}{2}-\kappa,\qquad k_2=\frac{k}{2}+\kappa.
\end{equation}
According to (30) and (34) $k_2>k_1$ and
\begin{equation}
{\rm e}^{ik_1N}={\rm e}^{ik_2N}=-1.
\end{equation}
Hence
\begin{eqnarray}
&&W=\frac{4}{N^2}\sum_{k_1<k_2}\dots=\frac{1}{2N^2}\sum_{k_1,k_2}\Big({\rm e}^{i[k_1(n_1-\tilde n_1)+k_2(n_2-\tilde n_2)]}+
{\rm e}^{i[k_2(n_1-\tilde n_1)+k_1(n_2-\tilde n_2)]}\nonumber\\
&&-{\rm e}^{i[k_1(n_1-\tilde n_2)+k_2(n_2-\tilde n_1)]}-{\rm e}^{i[k_2(n_1-\tilde n_2)+k_1(n_2-\tilde n_1)]}\Big)=\delta_{n_1\tilde n_1}\delta_{n_2\tilde n_2}-\delta_{n_1\tilde n_2}\delta_{n_2\tilde n_1}.
\end{eqnarray}
Now (38) follows from (A.12), (A.9) and (A.8).

\section{Evaluation of the form factor}
\setcounter{equation}{0}

Since
\begin{equation}
[{\bf S}^+(q),{\bf S}_n^-]=2\frac{{\rm e}^{-iqn}}{\sqrt{N}}{\bf S}^z_n,\qquad2{\bf S}^z_n|\emptyset\rangle=|\emptyset\rangle,\qquad\langle k-q|=\frac{1}{\sqrt{N}}\sum_n\langle\emptyset|{\bf S}_n^+{\rm e}^{i(q-k)n},
\end{equation}
one readily has
\begin{eqnarray}
&&\langle k-q|{\bf S}^+(q)|k,\kappa\rangle=\frac{2}{N\sqrt{N}}\sum_{1\leq n_1<n_2\leq N}\varphi_{n_2-n_1}(\kappa)\cos{\alpha(n_2-n_1)}\nonumber\\
&&=\frac{2}{N\sqrt{N}}\sum_{n=1}^{N-1}(N-n)\varphi_n(\kappa)\cos{\alpha n},
\end{eqnarray}
where $\varphi_n(\kappa)$ and $\alpha$ are given by (A.2) and (50). According to (11) and (50)
\begin{equation}
{\rm e}^{i\alpha N}={\rm e}^{\pm ikN/2},
\end{equation}
and hence (see (A.3))
\begin{equation}
\varphi_{N-n}(\kappa)\cos{\alpha(N-n)}=\varphi_n(\kappa)\cos{\alpha n}.
\end{equation}
So
\begin{equation}
\sum_{n=1}^{N-1}(N-n)\varphi_n(\kappa)\cos{\alpha n}=\sum_{n=1}^{N-1}(N-n)\varphi_{N-n}(\kappa)\cos{\alpha(N-n)}=\sum_{n=1}^{N-1}n\varphi_n(\kappa)\cos{\alpha n},
\end{equation}
and from (B.2) and (B.5) follows that
\begin{eqnarray}
&&\langle k-q|{\bf S}^+(q)|k,\kappa\rangle=\frac{1}{\sqrt{N}}\sum_{n=1}^{N-1}\varphi_n(\kappa)\cos{\alpha n}=\frac{1}{2N}\sum_{n=1}^{N-1}\Big({\rm e}^{i(\kappa+\alpha)n}+{\rm e}^{i(\kappa-\alpha)n}\nonumber\\
&&-{\rm e}^{i(\alpha-\kappa)n}-{\rm e}^{-i(\alpha+\kappa)n}\Big)=\frac{1}{2N}\Big(\frac{1+{\rm e}^{i(\kappa+\alpha)}}{1-{\rm e}^{i(\kappa+\alpha)}}+\frac{1+{\rm e}^{i(\kappa-\alpha)}}{1-{\rm e}^{i(\kappa-\alpha)}}
-\frac{1+{\rm e}^{i(\alpha-\kappa)}}{1-{\rm e}^{i(\alpha-\kappa)}}\nonumber\\
&&-\frac{1+{\rm e}^{-i(\kappa+\alpha)}}{1-{\rm e}^{-i(\kappa+\alpha)}}\Big)=\frac{1}{N}\Big(\frac{1+{\rm e}^{i(\kappa+\alpha)}}{1-{\rm e}^{i(\kappa+\alpha)}}+\frac{1+{\rm e}^{i(\kappa-\alpha)}}{1-{\rm e}^{i(\kappa-\alpha)}}\Big)=\frac{2i\sin{\kappa}}{N(\cos{\alpha}-\cos{\kappa})},
\end{eqnarray}
where we have put into account that according to (34) and (B.3)
\begin{equation}
{\rm e}^{i(\kappa\pm\alpha)N}=-1.
\end{equation}

\section{Evaluation of sums}
\setcounter{equation}{0}

First of all let us notice that according to (11) (37) and (50) one has
\begin{equation}
{\rm e}^{i\alpha_lN}=1,\qquad{\rm e}^{i\alpha_mN}=-1,\qquad{\rm e}^{i\kappa_{\lambda}N}=-1,\qquad{\rm e}^{i\kappa_{\mu}N}=1,\qquad N=2M.
\end{equation}
Hence
\begin{equation}
\sin{\alpha_m}\neq0,\qquad m=1,\dots,M.
\end{equation}

First of all we have to express the sums $\sum_{\lambda=1}^M$ and $\sum_{\mu=1}^{M-1}$ in (73) from the sums $\sum_{\lambda=1}^N$ and $\sum_{\mu=1}^N$. According to (35) and (36) for an arbitrary function $f(x)$ one has
\begin{eqnarray}
&&\sum_{\lambda=1}^Mf(\cos{\kappa_{\lambda}})=\frac{1}{2}\sum_{\lambda=1}^Nf(\cos{\kappa_{\lambda}}),\nonumber\\
&&\sum_{\mu=1}^{M-1}f(\cos{\kappa_{\mu}})=\frac{1}{2}\Big(\sum_{\mu=1}^Nf(\cos{\kappa_{\mu}})-f(1)-f(-1)\Big).
\end{eqnarray}

Also we shall need the identities
\begin{eqnarray}
&&\frac{1}{\cos{\kappa}-\cos{\alpha}}
=\frac{i}{\sin{\alpha}}\Big(\frac{1}{1-{\rm e}^{i(\kappa-\alpha)}}-\frac{1}{1-{\rm e}^{i(\kappa+\alpha)}}\Big),\qquad\cos{\alpha}\neq\pm1,\nonumber\\
&&\frac{1}{\cos{\kappa}-\cos{\alpha}}=\frac{2{\rm e}^{i\kappa}}{(\cos{\alpha}-{\rm e}^{i\kappa})^2}=2\Big(\frac{\cos{\alpha}}{(\cos{\alpha}-{\rm e}^{i\kappa})^2}-\frac{1}{\cos{\alpha}-{\rm e}^{i\kappa}}\Big)\nonumber\\
&&=2\cos{\alpha}\Big(\frac{1}{(1-\cos{\alpha}{\rm e}^{i\kappa})^2}-\frac{1}{1-\cos{\alpha}{\rm e}^{i\kappa}}\Big),
\qquad\cos{\alpha}=\pm1.
\end{eqnarray}

Let
\begin{equation}
{\tt S}(z)\equiv\sum_{j=1}^N\frac{1}{z-z_j},\qquad z_j^N=-1.
\end{equation}
Since $N=2M$ one has in (C.5) $(-z_j)^N=z_j^N=-1$. Hence additionally to (C.5)
\begin{equation}
\sum_{j=1}^N\frac{1}{z+z_j}={\tt S}(z),\qquad z_j^N=-1.
\end{equation}

Obviously
\begin{equation}
{\tt S}(z)=\frac{d}{dz}\log{\prod_{j=1}^N(z-z_j)}=\frac{d}{dz}\log{(z^N+1)}=\frac{Nz^{N-1}}{z^N+1}.
\end{equation}

Since according to (C.1)
\begin{equation}
{\rm e}^{i(\kappa_{\lambda}\pm\alpha_l)N}={\rm e}^{i(\kappa_{\mu}\pm\alpha_m)N}=-1,
\end{equation}
one readily has from (C.4)
\begin{eqnarray}
&&\sum_{\lambda=1}^N\frac{1}{\cos{\kappa_{\lambda}}-\cos{\alpha_l}}=\frac{i({\tt S}(1)-{\tt S}(1))}{\sin{\alpha_l}}(1-\delta_{\alpha_l,0}-\delta_{|\alpha_l|,\pi})\nonumber\\
&&-2\cos{\alpha_l}({\tt S}'(1)+{\tt S}(1))(\delta_{\alpha_l,0}+\delta_{|\alpha_l|,\pi})=-\frac{N^2}{2}\cos{\alpha_l}(\delta_{\alpha_l,0}+\delta_{|\alpha_l|,\pi}),\nonumber\\
&&\sum_{\mu=1}^N\frac{1}{\cos{\kappa_{\mu}}-\cos{\alpha_m}}=\frac{i({\tt S}(1)-{\tt S}(1))}{\sin{\alpha_m}}=0.
\end{eqnarray}

At the same time according to (C.4) one has
\begin{equation}
\frac{\sin^2{\alpha}}{(\cos{\kappa}-\cos{\alpha})^2}=-\Big(\frac{1}{1-{\rm e}^{i(\kappa-\alpha)}}-\frac{1}{1-{\rm e}^{i(\kappa+\alpha)}}\Big)^2.
\end{equation}
At $\sin{\alpha}\neq0$ this formula may be expanded to the form
\begin{equation}
\frac{\sin^2{\alpha}}{(\cos{\kappa}-\cos{\alpha})^2}=\frac{i}{\sin{\alpha}}\Big(\frac{{\rm e}^{-i\alpha}}{1-{\rm e}^{i(\kappa-\alpha)}}-\frac{{\rm e}^{i\alpha}}{1-{\rm e}^{i(\kappa+\alpha)}}\Big)
-\frac{1}{(1-{\rm e}^{i(\kappa-\alpha)})^2}-\frac{1}{(1-{\rm e}^{i(\kappa+\alpha)})^2}.
\end{equation}
Hence
\begin{eqnarray}
&&\sum_{\lambda=1}^N\frac{\sin^2{\alpha_l}}{(\cos{\kappa_{\lambda}}-\cos{\alpha_l})^2}=2({\tt S}(1)+{\tt S}'(1))(1-\delta_{\alpha_l,0}-\delta_{|\alpha_l|,\pi}),
=\frac{N^2}{2}(1-\delta_{\alpha_l,0}-\delta_{|\alpha_l|,\pi})\nonumber\\
&&\sum_{\mu=1}^N\frac{\sin^2{\alpha_m}}{(\cos{\kappa_{\mu}}-\cos{\alpha_m})^2}=2({\tt S}(1)+{\tt S}'(1))=\frac{N^2}{2}.
\end{eqnarray}
Using now (C.9) into (C.12) one readily gets
\begin{eqnarray}
&&\sum_{\lambda=1}^M\frac{\sin^2{\alpha_l}}{\cos{\kappa_{\lambda}}-\cos{\alpha_l}}
=-\frac{N^2}{4}\sin^2{\alpha_l}\cos{\alpha_l}(\delta_{\alpha_l,0}+\delta_{|\alpha_l|,\pi})=0,\nonumber\\
&&\sum_{\mu=1}^{M-1}\frac{\sin^2{\alpha_m}}{\cos{\kappa_{\mu}}-\cos{\alpha_m}}=\frac{\sin^2{\alpha_m}}{2}\Big(\frac{1}{1+\cos{\alpha_m}}-\frac{1}{1-\cos{\alpha_m}}\Big)=-\cos{\alpha_m},\nonumber\\
&&\sum_{\lambda=1}^M\Big(\frac{\sin^2{\alpha_l}}{(\cos{\kappa_{\lambda}}-\cos{\alpha_l})^2}-\frac{2\cos{\alpha_l}}{\cos{\kappa_{\lambda}}-\cos{\alpha_l}}\Big)=\frac{N^2}{4}\Big(1-\delta_{\alpha_l,0}
-\delta_{|\alpha_l|,\pi}\Big)\nonumber\\
&&+\frac{N^2}{2}\Big(\delta_{\alpha_l,0}+\delta_{|\alpha_l|,\pi}\Big)=\frac{N^2}{4}\Big(1+\delta_{\alpha_l,0}+\delta_{|\alpha_l|,\pi}\Big),\nonumber\\
&&\sum_{\mu=1}^{M-1}\Big(\frac{\sin^2{\alpha_m}}{(\cos{\kappa_{\mu}}-\cos{\alpha_m})^2}-\frac{2\cos{\alpha_m}}{\cos{\kappa_{\mu}}-\cos{\alpha_m}}\Big)=\frac{N^2}{4}
-\frac{1}{2}\Big(\frac{\sin^2{\alpha_m}}{(1-\cos{\alpha_m})^2}\nonumber\\
&&+\frac{\sin^2{\alpha_m}}{(1+\cos{\alpha_m})^2}\Big)+\frac{\cos{\alpha_m}}{1-\cos{\alpha_m}}-\frac{\cos{\alpha_m}}{1+\cos{\alpha_m}}=\frac{N^2}{4}-1.
\end{eqnarray}
The obtained system is equivalent to (73).

\end{document}